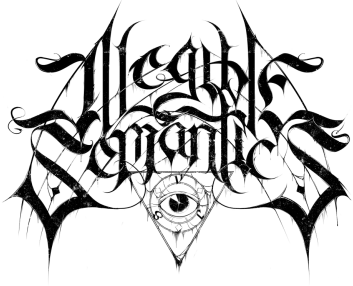

# Exploring the Design Space of Metal Logos

Gerrit J. Rijken🤘  Rene Cutura🤘  Frank Heyen🤘  Michael Sedlmair🤘  Michael Correll🤘
Tundra Toucan  University of Stuttgart  University of Stuttgart  University of Stuttgart  Tableau Research

Jason Dykes🤘  Noeska Smit🤘
City, University of London  University of Bergen

**Abstract**

The logos of metal bands can be by turns gaudy, uncouth, or nearly illegible. Yet, these logos *work*: they communicate sophisticated notions of genre and emotional affect. In this paper we use the design considerations of metal logos to explore the space of "illegible semantics": the ways that text can communicate information at the cost of readability, which is not always the most important objective. In this work, drawing on formative visualization theory, professional design expertise, and empirical assessments of a corpus of metal band logos, we describe a design space of metal logos and present a tool through which logo characteristics can be explored through visualization. We investigate ways in which logo designers imbue their text with meaning and consider opportunities and implications for visualization more widely.

## 1 Introduction

The ways that metal bands visually communicate is everything that data visualization is not, and yet metal stubbornly continues to succeed precisely where the visualization community has experienced failure after failure. Data visualization is often obsessed with "clarity," [13], an assumption that the very worst thing we could do for our audience is to not, as clearly and as minimally as possible, represent the dataset [7]. The logos of metal bands, by contrast, thrive in their unintelligibility, and in fact communicate sophisticated notions of genre and belonging and stance in ways orthogonal to how easy or difficult it might be to decode the "data" contained in the text. Similarly, data visualization has been self-described as the "unempathic art" [27], and struggles to create or deepen emotional connections [4, 14]. Yet, metal tightly integrates all sorts of death, despair, and derangement into its visual and musical aesthetics. We as a community have much to learn from the ways that bands and genres communicate with text and their complex *illegible semantics*.

In this paper we choose metal band logos, artifacts that fly in the face of all of our standard rules about design and legibility (and sometimes even good taste) as objects of study from an information visualization perspective. In short, our broad research question is: **what do metal logos visually communicate, and how do they manage to communicate so much despite their frequent illegibility**? In other words, how do visual rhetorics operate when the intent is not to be legible, but to perform a host of other important functions like communicating authority and belonging, emotion or intent, all while engaging with taboo or powerful subjects?

To address this question, we (a group consisting of visualization researchers from a variety of backgrounds as well as a professional graphic designer of metal band logos) present two parallel lines of inquiry: (1) An analysis of metal logos as visual form, resulting in a design space of 13 **Dimensions of Doom**: visual and typographic features, culminating in a design exercise where we see what implicit information we can communicate through iterations on a simple logo, and (2) **MetalVis**, a visual analysis tool that allows to explore a collection of 5,837 actual band logos based on their visual and semantic properties.

We use this paper to call for the visualization community to embrace the unintelligible, the macabre, and the hardcore: to think deeply, creatively, and continually about the relationships between data, text, and the messages that we deliver and hide, explicitly and subliminally, in our designs.

Code, tunes, data and other rocking materials are at: **illegiblesemantics.com**.

## 2 The Semiotics of Metal Logos

While visualization of musical data is an active area of research, as demonstrated in the comprehensive survey by Khulusi et al. [11], our focus is on how band *logos*, in concert with the musical acts they represent, perform communicative work. As such, we focus on the prior work around typography and visual semiotics. Van Leeuwen [25], shows how fonts and lettering can impart meaning: "A word can be changed into a 'warning' or a 'question' through typography and typographic signs alone [25]"— and enumerates features like weight, connectivity, and curvature that can change the way we view the text being written.

Comics present a useful example where lettering style and treatment is used to convey emotion, intent, and meaning. In *Comics and Sequential Art*, Eisner [9] shows an example where the dialogue in a comic drips blood. This was done, he writes, because "the effect of terror, implication of violence (blood) and anger brings the text into direct involvement." We can find further examples in the *DC Comics Guide to Coloring and Lettering Comics* [5], in which two stand out specifically given our metal focus: "bold, blocky letters for the thump of an impact" and "thin shaky letters for a ghostly wail", reminiscent of death and black metal respectively.

---

*🤘: These authors contributed equally to this work.

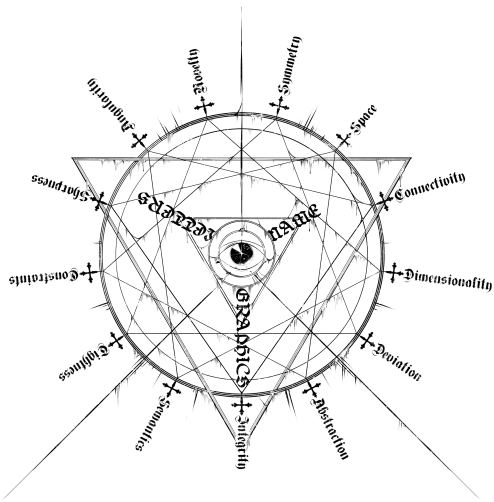

Figure 1: The 13 Dimensions of Doom visualized in a triskaidecagon.

Designers of metal band logos employ these techniques and more in their work: band logos have "typography that swirls like branches, drips like blood, and clings like spider webs" [22]. For example, Slayer's logo has "straight, sharp edges to reflect the tight and controlled nature of the music" [22]. As per founding member Dave Lombardo [15], this angular look was an intentional choice to reflect their name: "We were trying to figure out what the logo was going to look like and I said 'well, what is a murderer, what is a guy that kills: what does he do?' [pantomiming stabbing motions] So then what I did was I carved it."

Beyond communicating mood or themes, metal logo designers also use their designs to communicate group membership. Publisher Dayal Patterson [17] states "Logos in metal are often specific enough in their characteristics that they can tell the educated reader what subgenre is involved. In that sense they carry more information than merely the linguistic content. It's almost a visual code." Mark Riddick, author of *Logos from Hell* [20], a book-length compilation of metal logos, agrees [17]: "Although most bands tend to lean toward legibility for the sake of marketing, illegible logos also carry meaning and weight ... it can be assumed that the greater the illegibility the more extreme the band will be." The resulting coded information and nigh-illegible package can act as a way to repel outsiders: as per designer Tim Butler [22], "the point of these logos is like, unless you're in-the-know already, it's not for you."

The work that these visual conventions do is all the more visible in the breach, as with the example of the death metal band "Party Cannon," whose logo (a cheerful, Toys-R-Us style collection of brightly colored bubble letters) has been described as "the least death metal logo possible" [21]. The juxtaposition of the cheery logo with the logos of the other bands with which Party Cannon shares billings during tours or festivals (which tend to be black and white, angular, and severe) creates a memorable (and humorous) effect: as per bassist Chris Ryan [21], "every time it goes viral we definitely gain legitimate fans that listen to the music, attend shows, buy merch, and properly get what we're about."

## 3 Domain Analysis: Dimensions of Doom

As a team including a metal logo designer who has worked with numerous bands across genres over the last 20 years, 6 visualization researchers from academia and industry, and collectively 7 metal heads with many years of headbanging experience, we set out to uncover a set of dimensions that are specific to metal logo design and not covered by traditional typography style descriptions. The goal of our analysis was to determine a set of dimensions that are separable, groupable, and consistent. Our hypothesis is that metal band logos in selected sub-genres have characteristics that make them visually distinguishable from band logos in other sub-genres. We wanted to try to figure out the information carrying characteristics of metal logo design.

As a first step, Gerrit Rijken, the first author and professional designer, identified samples of logos that he considers associated with specific genres and described their visual characteristics. Subsequently, our team sketched examples, checked against existing band logos, discussed the characteristics and tested independence, meaning, and the terms adopted against those used in graphic design.

In parallel activity, the academics strapped on their headphones and looked to the literature, finding inspiration from semiotics legend Jacques Bertin [2], who talks of "letter" drawings, and takes a typographical perspective to ask "what are the independent variables ... that one can meet in the 'letter' and, possibly, in the word". He suggests two com-peting objectives in lettering. The first is **letter publicity**: characterized by "the freedom of imagination and the reduction of constraints": this is the flourish with which styles can be used to convey semantics indirectly, the unintelligibility of the logo that Eisner, Lombardo and Patterson celebrate and that we seek to explore. The second objective is **letter text** and is characterized by "increasing constraints in efforts to improve readability," i.e., the functional role that conveys information directly through language. Bertin identifies *onze lignes indépendantes en typographie* in his consideration of the subject [2] and describes letter orientation as "astonishingly spectacular", but "prohibited by technique and good practice," seemingly as the text objective takes priority over publicity. Along with the metal community, we challenge Bertin's view of "good practice" and playfully riff on his notion of "typographic amusements" by embracing practice in metal logo design, in which orientation is varied with a flourish and *styles*, to use Bertin's terminology, are varied widely but consistently. In the logo design space, immediate impact is essential and so letter publicity has a vital role, with letter text being less important than in many other contexts. This alternative approach draws attention to some important variables that are used to differentiate genres and imbue meaning at the cost of legibility.

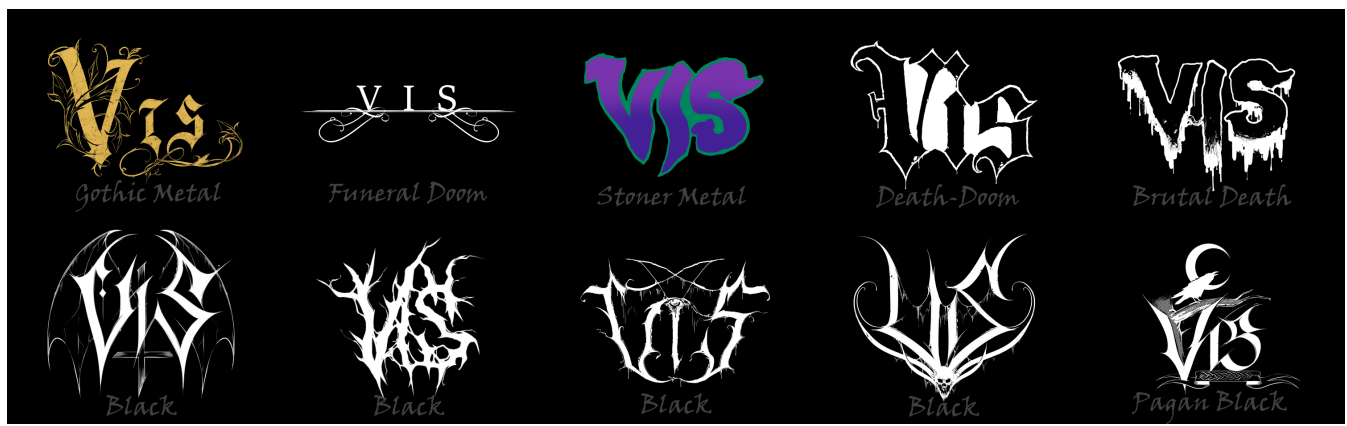

Figure 2: Logos in various styles representative for a subset of metal genres. From left to right, top row: gothic metal, funeral doom, stoner metal, old-school death, and brutal death. Bottom row: black (4x), and pagan black.

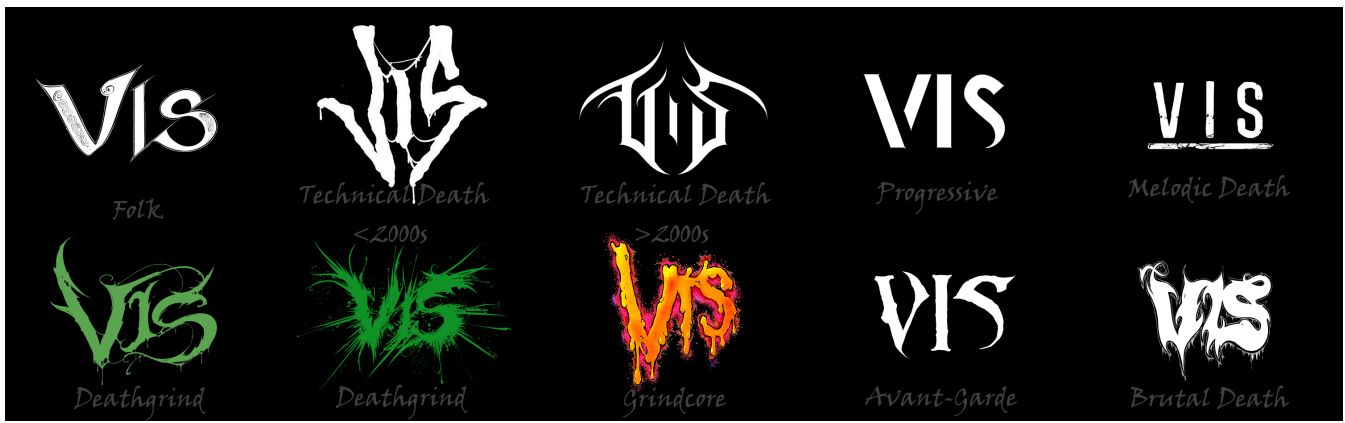

Figure 3: Logos in various styles representative for a subset of metal genres. Top row: folk, technical death (pre-2000s), technical death (post-2000s), progressive, and melodic death. Bottom row: deathgrind (2x), grindcore, avant-garde, and brutal death.

Through an intense iterative process, of cross-checking, discussion, sketching and even consideration of computer science (e.g. [24]), we agreed upon 10 characteristics of *style* used in the design of characters and logos in metal, and 3 characteristics of the graphics that often adorn them. The result is a set of 13 *Dimensions of Doom* which, when added to relevant characteristics from Bertin, might enable us to describe the broad superset of metal logos and differentiate between the genres that they represent. We summarize the dimensions that are key to metal logo design below, with detailed descriptions and visual samples available as supplementary materials. We do so in light of Bertin himself noting the importance of 'metal' and his penchant for it when identifying the "engraving effect" or "steel effect" as an important style [2], thereby alluding to his own alt credentials.

The 5 dimensions inspired by Bertin are: **Thickness** (how thick are the letters?), **Size** (how much variation is there in letter size?), **Texture** (how much texture is there in the letter rendering?), **Orientation** (how much variation is there from standard vertical letter orientation?), and **Color** (how many colors (hues) does the logo have in terms of color range?).

The 13 Dimensions of Doom are categorized in three groups. The first 5 dimensions are related to letter style: **Novelty** (how original is the font used in the letters of the logo?), **Angularity** (how angular are the outlines of the letters?), **Constraints** (how fixed are the angles of the letter segments?), **Sharpness** (how many prickly and sharp elements do the letters have?), and **Tightness** (how tight, clean, and precise are the letters?)

5 further dimensions relate to the entire logo: **Symmetry** (to what degree does the logo have vertical axis symmetry?), **Space** (how much negative space is left between letters?), **Connectivity** (how connected are the letters?), **Dimensionality** (how 3D does the logo look?), and **Deviation** (how much does the lettering deviate from the baseline?).

Finally, 3 dimensions relate to any additional graphics present: **Congruence** (how congruent and 'meaningful' are any graphical elements with respect to band name or genre?), **Abstraction** (how abstract are they?), and **Integrity** (how fully are the they integrated into the logo?).

### 3.1 Design Exercise

To explore and develop the dimensions, and see what they might tell us about metal logo design, Gerrit Rijken designed a variety of metal band logos drawing on his own experience and a wide range of iconic logos of bands in various genres of extreme metal. To ensure fair comparison, make the task manageable, and minimize confounding factors, we selected a small subset of letters, that collectively involve sharp edges, curved swirls, and opportunity for symmetry, for all genres: **VIS**. This resulted in a set of 30 uniquely designed metal band logos, as shown in Figures 2, 3, and 4. These were then developed in discussion with musicians and artists in the metal scene to ensure genre style accuracy and coverage. The list includes Jørn Inge Tunsberg (Hades Almighty, ex-Immortal), Dennis Jak (Dauthuz, Nembrionic, ex-Unlord), Frode Gaustad (Dominanz, ex-Thy Grief), Tjeerd Alwicher (Intero), Kristoffer Aalhus (Vinterbris, Funeral Void), and Korijn van Golen (Façade, Onheil), who confirmed that the visual styles of the logos adequately conveyed the intended genres. Due to the fractal, permeable, and protean nature of metal (sub)genres, our intent was not to fully cover the space, but to represent important "landmark" genres.

### 3.2 Designs and Dimensions

We then selected a subset of seven designs from the 30 **VIS** logos that covered a range of genres and styles and rated these according to the 13 Dimensions of Doom and the 5 Bertin dimensions. All authors rated each logo independently on a 5-point scale for each of the 18 dimensions (consult the supplemental materials for our responses). Through this process, we gained insights into which of the dimensions differentiated, which caused disagreement, whether the designer had a unique perspective and which dimensions were difficult to assess.

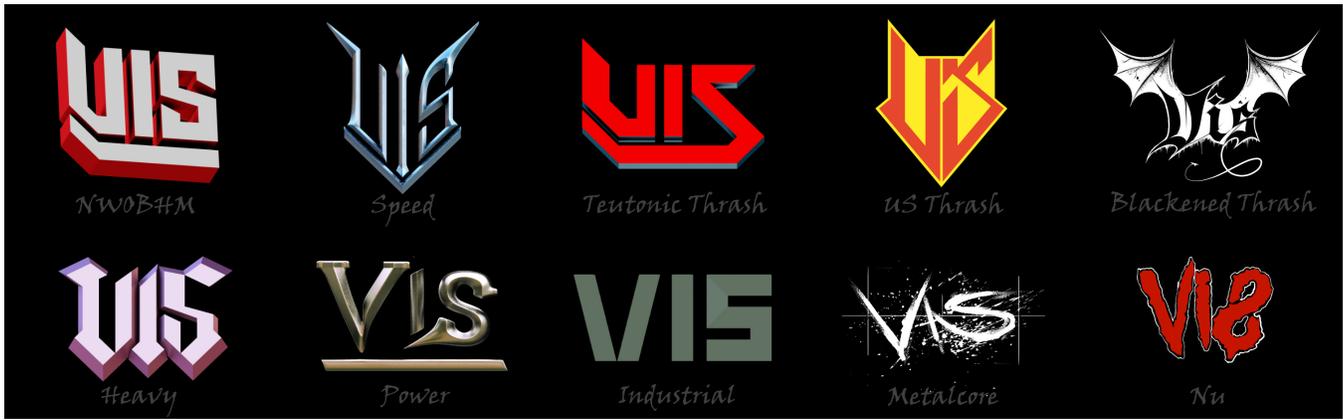

Figure 4: Logos in various styles representative for a subset of metal genres. From left to right, top row: new wave of British heavy metal, speed, teutonic thrash, US thrash, and blackened thrash. Bottom row: heavy, power, industrial, metalcore, and nu.

Briefly, we could separate some *specific logos* well with **Abstraction**, **Dimensionality**, **Space** and **Originality**, which have outliers. Certain *logo groups* can be discriminated well with **Sharpness**, **Integrity**, and **Symmetry**, which are somewhat bimodal. Discriminating between *all logos* works best with **Constraints**, **Texture** and **Angularity**, which are most regularly spaced. We detected some *disagreement between roles* – Gerrit's designer's eye versus the VIS researchers – on **Deviation**, **Size** and the **Additional Graphics** dimensions. These dimensions were the most challenging to assess, partly due to lack of clarity on what exactly constitutes an additional graphical element. Most *disagreement within the group* was found in assessing logo **Color**, **Tightness** and **Deviation**. In terms of *disagreement within the group across all dimensions*, our **VIS** logos were ranked

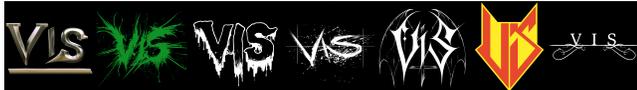

Another issue that arose is that the **Congruence** dimension requires an understanding of the intended genre or meaningful bandname, which was not part of this initial survey. We have sharpened the dimensions guide and definitions in light of these initial results.

## 4 Data-Driven Analysis: MetalVis

In parallel to this work with domain experts, we explored the metal logo design space in a data-driven manner.

To obtain a suitable corpus of existing metal band logos, we scraped the Encyclopaedia Metallum (metal-archives.com) for metal band logos and metadata, including band name and tags describing genre and lyrical themes. We gathered a total of 54,230 band logos and down-sampled this set by excluding bands that are unsigned or inactive, where the label has only one band, or where no lyrical themes were tagged. This resulted in a final set of 5,837 logos of bands considered to be active and somewhat successful.

Based on this data, we aimed to explore the logo design space by creating "*maps of metal*", in which band logos appear spatially close according to different definitions of similarity. In doing so, we follow previous cartographic attempts in this area, such as mapofmetal.com, where similarity is

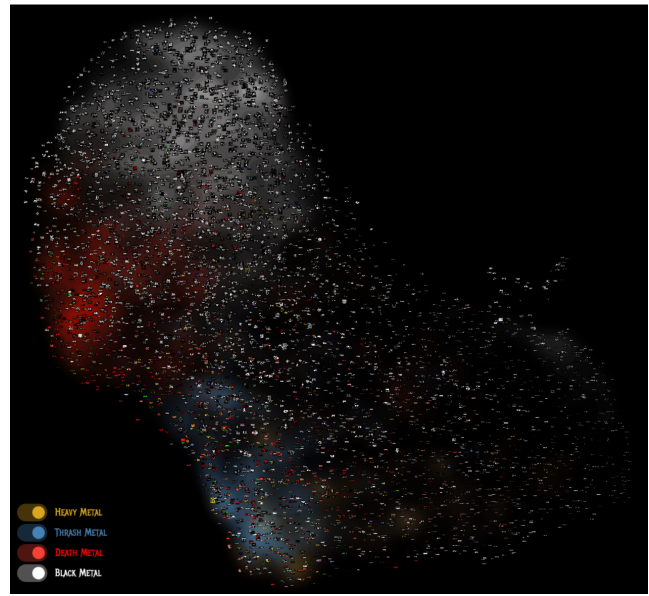

Figure 5: One of the maps of metal in *MetalVis*. Each point is a logo and the background color encodes the genre: white is black metal, red death metal, blue thrash metal, and gold is heavy metal.

based on genre and decade, and the map is manually crafted. Instead, we opted for an automatic layout and different similarity metrics characterizing the visual design of the band logos. Our interactive web-based tool, *MetalVis* (see Figure 5 and **illegiblesemantics.com**) is the result. It allows the logo design space to be explored through various approaches to defining the similarity between bands and logos:

Supervised modeling (logo/genre classifier)  To explore the latent visual space of metal logos, we trained a neural network classifier on our corpus, based on Inception v3 [23]. We used the logo images scaled to 160×160 pixels as input and predicted the 51 most common genres from the metadata as output variable. We then projected the last fully-connected 256D hidden layer to 2D with UMAP [16]. The results are shown in Figure 5. Some genres appear to form coherent clusters in the high-dimensional space. For example, black metal (white background) can be found at the top of the map. At the opposite end, we find both a thrash metal cluster (blue) as well as a heavy metal cluster (gold). Recall from Figure 2 bottom row and Figure 4 top row that these are indeed quite divergent in terms of visual style. This supports our hypothesis that the visual style of metal logos conveys information about the genre.

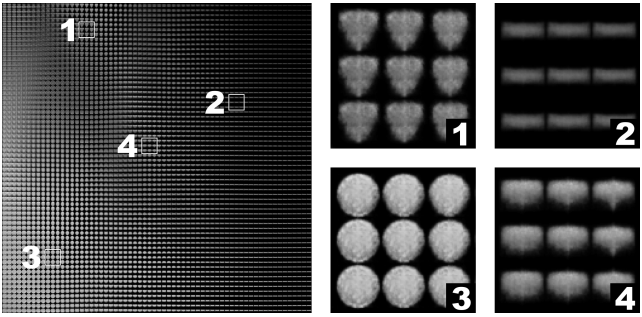

Figure 6: A regularly sampled plot of the VAE's latent space shows that the encoder separates logos based on their coarse shape. For example, there are regions of triangular (1), wide (2), disk-shaped (3), and sign-shaped (4) logos.

Unsupervised modeling (logo VAE) Since such a supervised classifier is biased toward the genre labels, we also tried a purely image-based unsupervised approach with a variational autoencoder (VAE). For this approach, we trained a rather simple VAE on 10,000 logos that were first re-scaled to 64×64 pixels and converted to greyscale. We then directly used the two-dimensional latent space that the encoder learned for the layout, by feeding all logos into the encoder and receiving two coordinates for each. The latent space contains regions that represent the logo's coarse shape (Figure 6).

Genre-based embedding Our map of logos in MetalVis can also be directly organized by the genres of bands (using UMAP's Sokal-Michener distance [16]). Doing so allows the user to "manually" explore the label designs in different genres. This approach allows us, for instance, to spot interesting outliers, where the visual style of the logo is incongruent with the genre, e.g., Cauldron, a heavy metal band with a death metal-style logo. It also allows us to explore cross-over bands and explore whether the visual styles of these logos are also mixing visual elements of their base genres, e.g., do blackened death metal bands have logos that feature both stereotypical black and death metal visual attributes?

Other features of MetalVis Besides these embeddings, we also tried out various other ways of organizing these maps of metals. For instance, we used the color of the logos and created a UMAP embedding based on the distance between color histograms. This approach allows us to learn about which genres tend to use which colors: red logos tend to be from death metal bands, while (somewhat counter-intuitively) black metal logos tend to be white. Standard interactions such as filtering, zooming and panning, and tooltips allow for an interactive exploration. In addition, all maps can be displayed using the original embedding with overlapping logos or a gridified version without overlaps, created with Hagrid [8]. Inquisitive metal-heads can check out the *MetalVis* tool at our companion website **illegiblesemantics.com**.

## 5 Discussion

While admittedly preliminary, we think both of our lines of inquiry, in concert with other methods like close reading [1], point to a useful way of interpreting and evaluating these rhetorical dimensions of visualization design. Rather than focusing exclusively on the data contained within them (and how easy or "efficient" [3] it is to extract these data), we should think about what sort of work our visualizations perform (or "what worlds they build" [6]) regardless of the specific datasets connected to them. Ease of information extraction is neither necessary nor sufficient to determine the success or failure of a visual design.

Just as metal logos communicate genre and belonging before the viewer has been able to "decode" their textual content, we think our exploration emphasizes the communicative work that visualizations perform before the viewer decodes the first data value. Kennedy et al. [10] point to this hidden "work" (such as the use of clean lines and two dimensional perspectives) that communicate senses of objectivity. Richards [19] claims that the manipulation of visuals to convey authority is an important part of how scientists work to make their arguments. What we get out of visualizations is influenced by titles [12], our preconceptions [26], and the visual metaphors we employ in our designs [28] in ways that are seemingly disconnected from the actual information represented in the visualization itself.

As with any analysis of meaning, we are colored by our own perspectives. We cannot, nor do we attempt to, speak for all metal designers, bands, or fans. Further complicating our work is that the notion of "genre" itself is an ever-shifting moving target. Bands can straddle multiple genres, which are themselves possessed of fuzzy boundary conditions subject to interpretation and disagreement. Bands themselves can alter their self-description and logos over time as they grow and change musically.

In the future, we hope to both *extend*, *refine*, and *unify* our parallel explorations. For instance, variability and ambiguity in our interpretation of dimensions point to a need to iterate on our dimensions and evaluate them more formally. We also believe that including additional features (such as time, geography, or our Dimensions of Doom) in our quantitative models could allow us to tell a more robust and holistic story of how visual style creates meaning in band logos.

Just as simple visualizations like scatterplots can act as metaphorical "fruit flies" for visualization research [18], we are convinced that further study of metal logos, with their heterogeneity and polysemic structure, deep cultural complexity and gloriously illegible semantics, could be useful for testing or developing models around open visualization research problems in visual rhetoric, affect, and visual similarity.

### 5.1 Conclusion

We focus on metal logos as an object of study. Not just because they 🤘rule🤘, but because they manage to accomplish what so much mainstream data visualization struggles to achieve: making us energized or emotional, uncomfortable or afraid. And they accomplish this without getting hung up on "clarity" or "readability," "data-ink" or "discoverability," but rather by embracing and celebrating letter publicity at the cost of letter text. Our preliminary analysis is beginning help us understand this design space and offers plenty of scope for developing our knowledge of metal logos, the messages that they impart, genres that they represent and what this might mean for visualization. Just as metal culture, on the fringes of society, can be a powerful commentary on our society, so too can metal logos, at the fringes of graphic design, serve as commentary on the visualization community.

**Acknowledgments**

The authors wish to thank the entire metal scene for inspiration, particularly those listed in the eclectic *Illegible Semantics playlist*. Partially funded by the Deutsche Forschungsgemeinschaft (DFG) – Project-ID 251654672 – TRR 161, and by the Cyber Valley Research Fund – Project InstruData.